\documentclass[12pt,preprint]{aastex}

\usepackage{natbib}
\usepackage{times}
\usepackage{sidecap}

\shortauthors{Lionello et al.}
\shorttitle{Validating a Time-Dependent Wave-Turbulence-Driven Model of the Solar Wind}

\begin{document}


\title{Validating a Time-Dependent Turbulence-Driven Model of the Solar Wind}
\author{Roberto Lionello}
\affil{Predictive Science, Inc.,  9990 Mesa Rim Rd., Ste. 170, San Diego, CA
92121-3933}
\email{lionel@predsci.com}
\author{Marco Velli} 
\affil{Jet Propulsion Laboratory, California Institute of Technology,
4800 Oak Grove Dr., Pasadena, CA 91109}
\email{mvelli@mail.jpl.nasa.gov}
\author{Cooper Downs, Jon  A.\ Linker, 
Zoran Miki\'c}
\affil{Predictive Science, Inc.,  9990 Mesa Rim Rd., Ste. 170, San Diego, CA
92121-3933}
\email{\{cdowns,linker,mikic\}@predsci.com}
\author{Andrea Verdini}
\affil{Observatoire Royale de Belgique, 3 Avenue Circularie, 1180, Bruxelles,
Belgium}
\email{verdini@oma.be}


\begin{abstract}
Although the mechanisms responsible for heating the Sun's corona and
accelerating the solar wind are still being actively
 investigated, it is largely
accepted that photospheric motions provide the energy source and that
the magnetic field must play a key role in the process. \citet{2010ApJ...708L.116V}
presented a model for heating and accelerating the solar wind based on
the turbulent dissipation of Alfv\'en waves.  We first use a time-dependent
model of the solar wind to reproduce one of 
 \citeauthor{2010ApJ...708L.116V}'s solutions; then
we  extend its application to the case when the
energy equation includes thermal conduction and radiation losses, and the
upper chromosphere is part of the computational domain. Using this model,
we explore parameter space and describe the characteristics of a fast-solar-wind solution.
We  discuss how  this formulation may be applied to a 3D MHD model  of the corona and
solar wind \citep{2009ApJ...690..902L}.
\end{abstract}


\section{INTRODUCTION}
The identification of the 
physical processes responsible for the heating of the solar corona
and the acceleration of the solar wind still represents 
an unsolved problem in solar physics. However, there is a general consensus 
that photospheric motions provide the energy source and that the
magnetic field must play a key role in the process. 
Since  it is clear that the measured speeds of fast streams require
an extended heating deposition
\citep{1977ARA&A..15..363W,1980JGR....85.4665H,1988ApJ...325..442W},
previous
one-dimensional (1D) models  generally relied on a parametric heating
function exponentially decaying with height
\citep{1982ApJ...259..779H,1982ApJ...259..767H,1988ApJ...325..442W,
1995JGR...10021577H,1995GeoRL..22.1465R,1997ApJ...482..498H}.
At the same time, several investigations were based 
on low-frequency broadband fluctuations on magnetohydrodynamic scales 
as the mechanism that heats and accelerates the 
solar wind
\citep{1968ApJ...153..371C,1971JGR....76.3534B,1986JGR....91.4111H,
1988JGR....93.9547H,1994AdSpR..14..123V,1999ApJ...523L..93M,2007ApJ...662..669V,2012ApJ...745...35Z}.

Connecting the macroscopic
heating of the plasma and acceleration of the wind, as
 formulated in  coronal
 and inner heliospheric MHD models, with the underlying physical 
mechanisms is complicated due to the temporal and spatial
dynamic ranges involved.
In  the past decade,
turbulent dissipation mechanisms have been
 progressively incorporated with various degrees of
self-consistency into 1D models   of the  solar wind
\citep{2005ApJ...632L..49S,2005ApJS..156..265C,2007ApJS..171..520C,2010ApJ...710..676C,
2010ApJ...708L.116V,2011ApJ...743..197C}. 
There are also efforts to replace
empirical heating functions in three-dimensional (3D) MHD models with 
some form of  turbulence dissipation mechanism. 
This is particularly challenging because 
such 3D models would have to resolve time-scales extending from a
millisecond (dissipative time-scale in the solar corona) up to many days (large scale solar wind stream structure). \citet{2010ApJ...725.1373V}
introduced, beside the acceleration of the solar wind through Alfv\'en waves,
the  heating of the protons by 
Kolmogorov dissipation in open field-line regions.
\citet{2011ApJ...727...84U} developed a large-scale MHD heliospheric model 
with small-scale transport equations for the turbulence
energy, normalized cross helicity, and correlation scale, applicable where the solar wind is already supersonic and superalfv\'enic. 
The model of
\citet{2013AIPC.1539...30L} included  the effect of outwardly propagating Alfv\'enic
turbulence in the solar wind  and a phenomenological term to describe
nonlinear interactions associated with wave reflection by density gradients in the
chromosphere and corona. \citet{2013ApJ...764...23S} and 
introduced
Alfv\'en wave turbulence, assuming that
this turbulence and its nonlinear dissipation are the only momentum
and energy source for heating the coronal plasma and driving the solar wind
{\citep[see][for additional details]{2013arXiv1311.4093V}. }

The present work illustrates the integration of  the turbulence
dissipation heating and acceleration
 mechanism of \citet{2010ApJ...708L.116V} in a time-dependent,
1D, hydrodynamic (HD) model of the solar wind, which includes also thermal conduction
and radiation losses. 
In this early, explorative phase,
 it is expedient to conduct an investigation using 1D models so that we may later
apply our gained experience to  3D models. 
The model of \citet{2010ApJ...708L.116V} employs
strong turbulence closure to treat nonlinear effects, and  does not rely on 
electron heat conduction for      radial energy transport, but rather 
computes the  internal energy associated with protons only. 
After presenting the characteristics of our {turbulence-driven
 HD} model,
we show how it can match one of the solutions derived by
\citet{2010ApJ...708L.116V} without thermal conduction and radiation losses.
Subsequently,
we extend the application of our model to  include    transport
mechanisms 
in the energy equation, which are necessary to reproduce plasma emission    
in agreement with observations \citep{2009ApJ...690..902L}.
Our exploration of parameter space 
yields solutions  compatible with the solar wind properties obtained
from \textit{in situ} measurements and observations. In the future we 
 plan to introduce
this formulation into the 3D MHD thermodynamic  model of the solar corona
and solar wind of \citet{2009ApJ...690..902L}.

This paper is organized as follows:
the equations
and the solution technique are described
 in Sec.~\ref{sec-model}. In
Sec.~\ref{sec-results} we present  a solution in the configuration
of \citet{2010ApJ...708L.116V}, we conduct a parameter study that
 includes thermal conduction and radiative losses, and we describe 
 the details of
one of the solutions.
We conclude with a discussion.


\section{MODEL DESCRIPTION}
\label{sec-model}
\citet{2010ApJ...708L.116V} solved a steady-state, 1D model of the solar wind
along an expanding flux tube that included
heating and acceleration from turbulent dissipation.
We solve a more general  set of time-dependent, 1D HD equations 
along an open magnetic field line that can be reduced to the model of
\citet{2010ApJ...708L.116V} as a special case:
\begin{eqnarray}
\frac{\partial \rho}{\partial t}
  &=&  
-  \frac{1}{A}\frac{\partial}{\partial s}\left ( A U \rho \right ) , \label{eq-rho}
  \\
\rho  \frac{\partial U}{\partial t} &=& 
- \rho  U\frac{\partial U}{\partial s} - \frac{\partial }{\partial s}(p+p_w) + g_s \rho
+ \mathrm{R}_s  
 + \frac{1}{s^2}\frac{\partial}{\partial s} 
\left (s^2 \nu  \rho \frac{\partial U}{\partial s} \right ) , \label{eq-mom}
 \\
\frac{\partial T}{\partial t}
 &=& -U\frac{\partial T}{\partial s}
  -(\gamma -1)\left ( T {\frac{1}{A}\frac{\partial}{\partial s}A} U 
 -\frac{m_p}{2 k \rho}
   \left( {\frac{1}{A}\frac{\partial}{\partial s}A} q 
    -  n_en_p{Q(T)}+{H}\right) \right),
\label{eq-T} 
\end{eqnarray}
where $s \geq  R_\odot$ is the distance along a  magnetic field line, which may differ from the
radial distance from the surface, $r$; 
$\rho$, $U$, $p=2 n k T$, and $T$ are the plasma mass density, velocity,
pressure, and temperature ($k$ is Boltzmann constant and $n$ the number
density);  $g_s=g_0 R_\sun^2 \mathbf{\hat{b}\cdot \hat{r}}/r^2$
is the gravitational acceleration along the direction of the field line
($\mathbf{\hat{b}}$), and $\nu$ is the  kinematic viscosity. $A(s)=1/B(s)$ is
the area factor, i.e., the inverse of the magnetic field magnitude $B(s)$, along the
field line. $R_\mathrm{s}$ is the field aligned component of the 
vector divergence of the MHD Reynolds stress,
$\mathbf{R}=(\delta \mathbf{b}  \delta \mathbf{b}/ 4 \pi - \rho \delta
\mathbf{u}  \delta \mathbf{u} )$, where $ \delta\mathbf{b},  \delta\mathbf{u}$ are
respectively the
fluctuations of the magnetic field, $\mathbf{B}=B(s) \mathbf{\hat{b}} +\delta  \mathbf{b}$ and velocity
$\mathbf{u}=U(s) \mathbf{\hat{b}} + \delta \mathbf{u}$, with  $\mathbf{\hat{b}} \cdot \delta
\mathbf{b} = 0= \mathbf{\hat{b}} \cdot \delta \mathbf{u}          $. The
wave pressure term is
$ p_w = {\delta \mathbf{b}^2}/{8 \pi}  $. {This form of the equations 
is correct for incompressible fluctuations in planes orthogonal to the density gradients,
and though written somewhat differently corresponds to the equation originally derived in
\citet{1980JGR....85.1311H} 
and discussed in detail, from the linear point of view, in \citet{1993A&A...270..304V}. }
Equation (\ref{eq-T}) contains the
radiation loss function $Q(T)$ as in \citet{1986ApJ...308..975A},
$n_e$ and $n_p$ are the electron and proton
number
density (which are equal    for a hydrogen plasma), $\gamma=5/3$ is
the polytropic index, ${q}$ is  the heat flux. A
collisional (Spitzer's law) or collisionless \citep{1978RvGSP..16..689H}
formulation is used according to the radial distance,
\begin{equation}
{q}=\left \{ \begin{array}{ll}
\displaystyle
 - \kappa_0 T^{5/2} 
      \frac{\partial T}{\partial s }  & \mbox{if $R_\odot \leq r \lesssim 10R_\odot$} \\
     \alpha   n_ekT U & \mbox{if $r \gtrsim 10R_\odot$}
             \end{array} \right . ,
\end{equation}
where $\kappa_0=9\times10^{-7}$~erg~K$^{-7/2}$~cm$^{-1}$~s$^{-1}$ and  $\alpha$
is a parameter, which is set to 1.
The transition between
the two forms  occurs
smoothly {at a distance of $10R_\odot$ from the Sun}
\citep{1999PhPl....6.2217M},
{approximately where the radial
electron mean free path
becomes equal to the radial ``trapping distance"
\citep{1976JGR....81.1649H,1978RvGSP..16..689H}.}
To write an expression for
the heating function per unit volume $H$, which depends  on the perturbations
\citep{1938RSPSA.164..192D,2004GeoRL..3112803M}, it is more convenient to 
use the Elsasser variables
$\mathbf{z}_\pm=\delta \mathbf{u}
\mp \delta \mathbf{b}/\sqrt{4 \pi \rho} $ \citep{2001ApJ...548..482D}.
$\mathbf{z}_+$ ($\mathbf{z}_-$) represents an outward (inward) propagating
perturbation along a radially outward 
 magnetic field line. We shall assume that the 
actual direction of $\mathbf{z}_\pm$ is not important, as long as it is in
the plane perpendicular to $\mathbf{\hat{b}} $ and that only low-frequency
perturbations are relevant for the heating and acceleration of the solar wind. Hence, 
we write the incompressible, volumetric heating  
\citep{1980PhRvL..45..144D,1983A&A...126...51G,1995PhFl....7.2886H,2004GeoRL..3112803M}
in terms of  
$z_\pm$, the Fourier components  in the zero-frequency limit, as
\begin{equation}
\label{eq-scale}
H=  \rho  \frac{|z_-| z_+^2+ |z_+| z_-^2}{ 4 \lambda_\perp                 }.
\end{equation}
Here $\lambda_\perp$ is the correlation
scale of the turbulence, which we assume to be related to its value
at the solar surface, $\lambda_\odot$, according to the following formula:
\begin{equation}
\lambda_\perp=\lambda_\odot \sqrt{\frac{A}{A_\odot}},
\end{equation}
where  $A_\odot$ is the flux tube area at        $r=R_\odot$.

The evolution of $z_\pm$ is described by the following equations
\citep{2007ApJ...662..669V}:
\begin{eqnarray}
\frac{\partial z_\pm}{\partial t}
&=&
-[U\pm V_a] \frac{\partial z_\pm}{\partial s}+
R_1^\pm z_\pm +R_2^\pm z_\mp - \frac{|z_\mp| z_\pm}{2\lambda_\perp
}, \label{eq-zpm}
\\
R_1^\pm&=&-\frac{1}{2}[U\mp V_a]\left ( \frac{\partial \log V_a}{\partial s}
+\frac{\partial \log A}{\partial s} \right ),  \\
R_2^\pm&=&\frac{1}{2}[U\mp V_a]\frac{\partial \log V_a}{\partial s},
\end{eqnarray}
where $V_a(s)=B/\sqrt{4 \pi \rho}$ is the Alfv\'en speed along the field line.
$R_1^\pm $ and $R_2^\pm$ are respectively  the WKB and reflection terms related to the large scale
gradients. 
{ This equation is time-dependent so a clarification is in order.
Generally speaking, the linear terms in this equation are correct for Alfv\'en
waves of any frequency. However the nonlinear damping term, which is consistent
with conservation of energy together with the form (\ref{eq-scale})
for the heating, is really a dimensional estimation of the full non-linear cascade rate (including wave-vector components in the perpendicular directions).
For any energy spectrum which is a decreasing power -law with frequency, it is the energy in the lowest frequencies which dominates 
the cascade rate, and one can then remove averages involved in obtaining the
cascade rate and heating function as long as the fluctuations have frequencies
which are small. If one had finite frequency fluctuations, a little more
caution would be required, as the heating terms in  Eq.~(\ref{eq-scale})
would involve time-averages of the quadratic fluctuating quantities and the absolute value would be replaced by the square root
of the average squared fluctuation, and also in Eq.~(\ref{eq-zpm}) one should
replace the absolute value of the opposite Elsasser variable again with the square root
of the average squared quantity, as discussed in \citet{2010ApJ...708L.116V}.}

{The form of scaling for $\lambda_\perp$ in Eq.~(\ref{eq-scale})
  is a common assumption
in the literature. We use it instead of more sophisticated expressions
\citep[e.g.][]{2008JGRA..113.8105B} because, in future 3D applications,
 it will not require us to integrate along magnetic field lines. 
Furthermore, more complicated formulations introduce only a 
second-order correction in the evolution of $z_\pm$ in the average of the
opposite Elsasser variable would appear in the nonlinear transport term 
Eq.~(\ref{eq-zpm}) rather than more physics.  }
We can now express both $p_w$ and $\mathrm{R}_s$ in terms of   $z_\pm$ as
\citep{2011ApJ...727...84U,2012ApJ...754...40U}
\begin{eqnarray}
p_w&=& \rho \frac{ (z_- -z_+)^2}{8}, \\
\mathrm{R}_s&=& \rho \frac{z_+ z_- }{2}  \frac{\partial \log A}{\partial s} .
\end{eqnarray}

The equations to be solved are quasi-hyperbolic \citep[i.e,  some of the 
eigenvalues coincide in certain limits, e.g.][p.~413]{2010adma.book.....G},
 so at the Sun, where
the solar wind is subsonic, only density and temperature can be specified, while the velocity is determined determined by solving
the gas characteristic equations. At the upper radial boundary, which is placed
beyond all critical points,  the characteristic equations are used as well. The $z_\pm$ equations are also hyperbolic, and the
solar wind is sub-Alfv\'enic at the inner boundary, so we can impose only the amplitude of
the outward-propagating (from the Sun) wave.
Nonuniform meshes can be
specified to concentrate the resolution in the
transition region, where large gradients are present.  The use of
 a semi-implicit
treatment of the Alfv\'en and magnetosonic waves
in the momentum equation, Eq.~(\ref{eq-mom}), allows us to specify time steps larger than
the CFL limit
\citep{1998JCoPh.140....1L,1999JCoPh.152..346L}.
The special treatment of the thermal conduction and radiation loss function,
as explained in \citet{2009ApJ...690..902L}, is used to lower the
gradients in the transition region without significantly affecting the coronal
solution.

We express the area factor $A(s)$  in terms of $f(s)=A(s)/s^2 $, where the expansion
factor \citep{1976SoPh...49...43K} is
\begin{eqnarray}
f(s)&=&\frac{f_\mathrm{max}+f_1 \exp[-(s-r_f)/\sigma] }{1+\exp[-(s-r_f)/\sigma]},
 \\
f_1&=& 1 - (f_\mathrm{max}-1)\exp[(1-r_f)/\sigma]. \nonumber
\end{eqnarray}
In the present work we consider a radial field line (i.e., $s=r$) with
the same parameters as those 
 used by \citet{2010ApJ...708L.116V}, namely 
$f_\mathrm{max}=12.5$, $r_f=1.31~R_\odot$, and $\sigma=0.51~R_\odot$. The  magnetic
field magnitude at 1 AU is $B=3\times10^{-5}~G$.

\section{RESULTS}
\label{sec-results}
Here we present applications of the model described in Sec.~\ref{sec-model}. We
first perform a calculation  without thermal conduction and radiative losses
to recover a solution as in  \citet{2010ApJ...708L.116V}. Then we extend the
model by including energy transport mechanisms in the energy equation, we
complete a parameter study, and show
the characteristics of a fast solar wind solution.
\subsection{A Solar Wind Solution After \citet{2010ApJ...708L.116V}}
\label{sec-verdini}
\begin{figure}
\includegraphics[width=\textwidth]{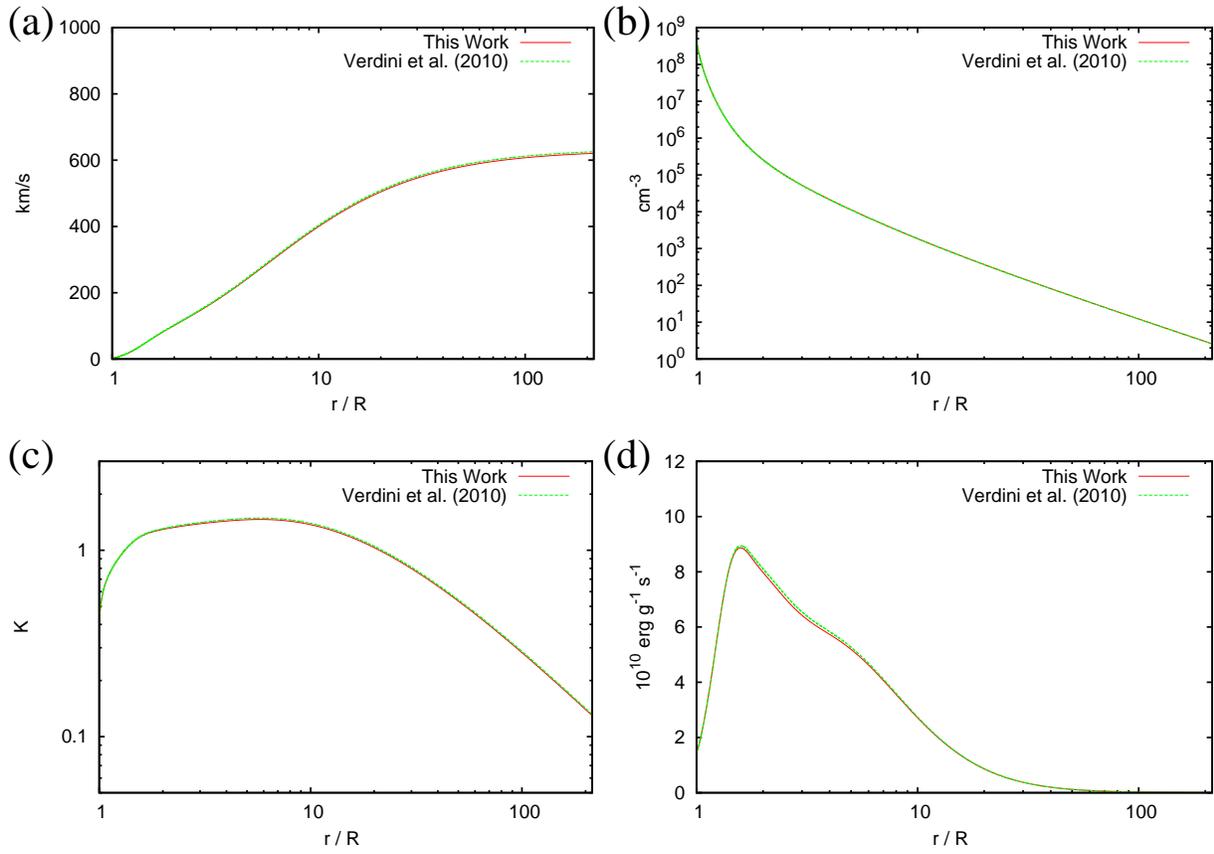}
\caption{Comparison between the solution obtained with the model described in
this work and that obtained with the model of \citet{2010ApJ...708L.116V}: (a) wind speed; (b)
density; (c) temperature; (d) total heating per unit mass. The discrepancies
are less than $1\%$. }
\label{fig-verdini}
\end{figure}

We now apply the model described in the previous section to obtain a solar wind solution
with the same parameters used in
\citet{2010ApJ...708L.116V}.
 The model of
\citet{2010ApJ...708L.116V} has neither   thermal conduction  nor radiation
losses but adds instead a small compressional heating term in the lower
corona:
\begin{equation}
\frac{H_c}{\rho}=3\times 10^{10}  
\exp \left ( - \frac{\left (r/R_\odot -x_g \right )^2 }{2 \sigma_g^ 2}
 \right )~\mathrm{cm}^2\mathrm{s}^{-3},
\label{eq-comp}
\end{equation}
with $x_g=1.3$, $\sigma_g=0.5$. 
The boundary conditions at the base          are $T_\odot=4 \times
10^5~\mathrm{K}$, $n_\odot=5 \times 10^8~\mathrm{cm}^{-3}$. Among the different
solutions calculated by \citet{2010ApJ...708L.116V}, we consider the one that
has 
$\lambda_\odot=0.015~R_\odot$, $z_+^\odot=37.24~\mathrm{km/s}$ at the solar
surface, which
yields a velocity perturbation $\delta u_\odot= 20~\mathrm{km/s}$.
A nonuniform mesh is employed with 1705 points and $\Delta s$ ranging from
$7.8\times 10^{-4}~R_\odot$ at the solar surface to $1.2~R_\odot$ at 1 AU. A
small kinematic viscosity, such that {ratio of}  the associated dissipation
time with the propagation time of Alfv\'en waves  is $\tau_\nu/\tau_A=5000$, 
is added to damp  unresolved
scales below grid resolution \citep{2009ApJ...690..902L}. Contrary
to the model \citet{2010ApJ...708L.116V},  by default our model
accounts for thermal conduction and radiation losses.
 For numerical reasons, rather then immediately removing the
said   terms from the energy equation,
we start the simulation with a guess solution
and advance Eqs.~(\ref{eq-rho}-\ref{eq-T} and \ref{eq-zpm}) (i.e., including
also
energy transport) for the first 80
hours. Then, for the successive 40 hours, we gradually decrease to zero the thermal
conduction and radiation contributions in Eq.~(\ref{eq-T}) by multiplying them
with a linear function $f(t)$ such that $f(80~\mathrm{hrs})=1$ and
$f(120~\mathrm{hrs})=0$. Finally, having turned-off the energy transport
operators, we let the system
relax for 200 more hours. In Fig.~\ref{fig-verdini} we show a comparison
between the solutions obtained with the present model and that of 
\citet{2010ApJ...708L.116V}  for wind speed, density, temperature, and total heating
per unit mass.
The two models appear to agree within less than $1 \%$.

\subsection{Solar Wind Solutions With Energy Transport}
\label{sec-thermo}
\begin{figure}
\includegraphics[width=\textwidth]{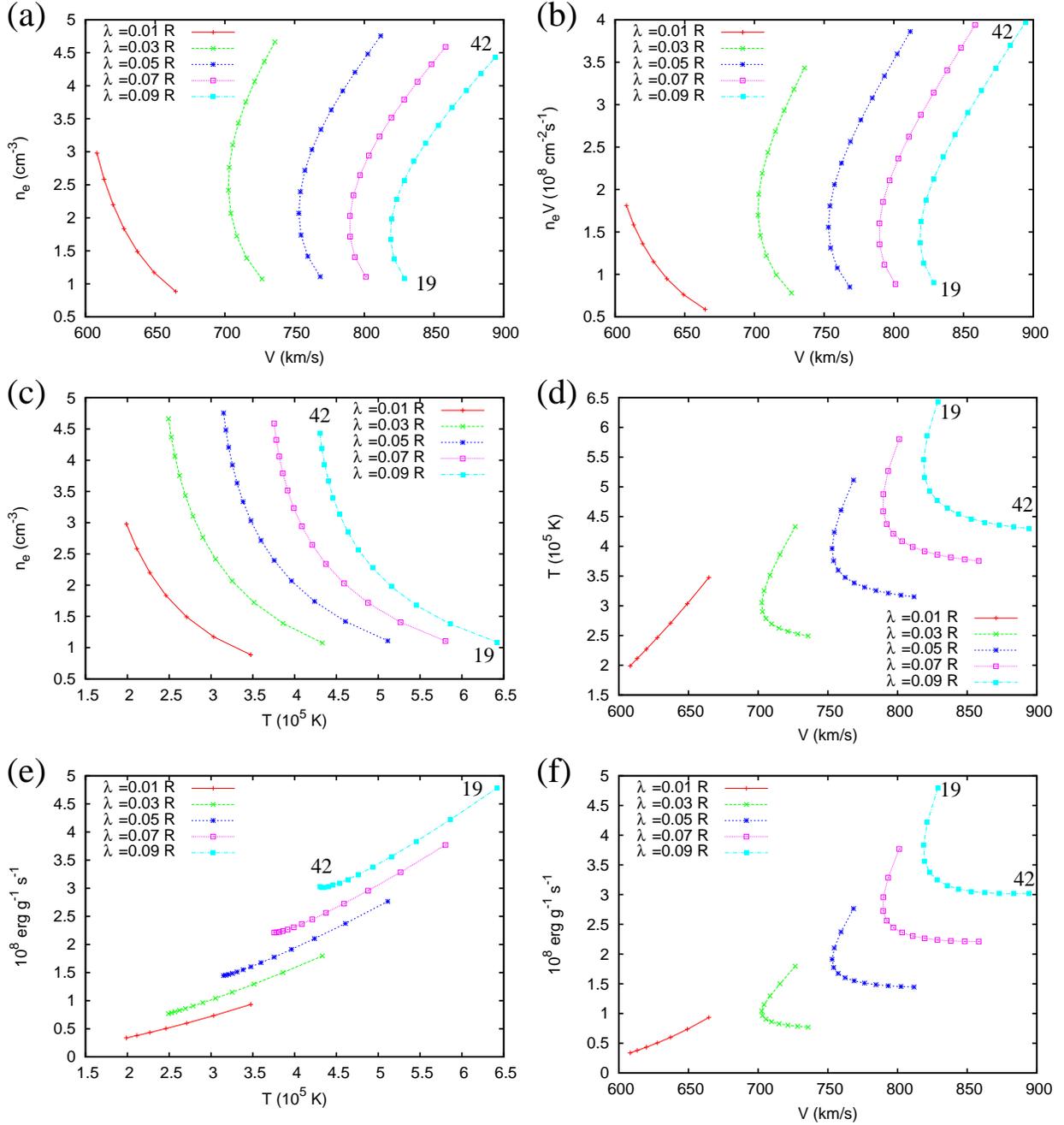}
\caption{Values at 1 AU of wind speed, density, mass flux,  temperature,
and heating per unit mass
for different combinations of $z_+^\odot$ (amplitude of the perturbation at the
base) and $\lambda_\odot$ (correlation scale of the turbulence). For each curve
of constant $\lambda_\odot$,  $z_+^\odot$ is varied uniformly between 19 and 42 km/s
(except the case with
$\lambda_\odot=0.01~R_\odot$ and $\lambda_\odot=0.03~R_\odot$ for which
no steady-solution is found at the highest values of $z_+^\odot$): (a) 
density vs.\ wind speed; (b) mass flux  vs.\ wind speed; (c) density vs.\
temperature; (d) temperature vs.\  wind speed; (e) heating per unit mass vs.\
temperature; (f) heating per unit mass vs.\ wind speed.
}
\label{fig-1AU}
\end{figure}
Having shown that we can reproduce a solution of \citet{2010ApJ...708L.116V} 
without thermal conduction and radiation losses,
we now reintroduce them into Eq.~(\ref{eq-T}) to complete a parameter
space study. We also use different  values
 of density and temperature at the base of the domain of $n_0=2\times
10^{12}~\mathrm{cm^{-3}}$ and $T_0=20{,}000~\mathrm{K}$ respectively, to
include the transition region and upper chromosphere in the calculation.
The use of these boundary conditions at the base of the chromosphere
is described by \citet{2005ApJ...621.1098M} and \citet{2013ApJ...773...94M}. 
{
The exact value of the density is not important, as long as it is large
enough to form a temperature plateau at the top of the
chromosphere. As shown in \citet{2009ApJ...690..902L},
while using smaller values is possible, there is a risk of evaporation
for the chromosphere. On the other hand, using larger values will slightly
increase the thickness of the plateau, without significantly affecting
the properties of the corona.
}
A  technique that
artificially broadens the
transition region, while maintaining accuracy in
the corona is likewise employed  \citep{2009ApJ...690..902L}.
The compressional heating term $H_c$ of Eq.~(\ref{eq-comp}) is set to zero.
For the same field line described in Sec.~\ref{sec-model}, we try to calculate
steady-state solar wind solutions with varying values of $z_+^\odot$ 
and $\lambda_\odot$ at the
base. We consider 13 different values of  $z_+^\odot$ equally spaced
between 
 $19~\mathrm{km/s} \lesssim z_+^\odot \lesssim 42~\mathrm{km/s}$, the interval
between each value being $\Delta
z_+^\odot \simeq 1.9~\mathrm{km/s}$. For $\lambda_\odot$ we
choose 5 values   within 
 $0.01~R_\odot\leq \lambda_\odot \leq 0.09~R_\odot$,
 with an interval $\Delta \lambda_\odot=0.02~R_\odot$.

We are not able to obtain steady-state solution for all the values selected.
 In fact, when $\lambda_\odot=0.01~R_\odot$, non-oscillatory
solutions are possible only for $19~\mathrm{km/s} \lesssim z_+^\odot \lesssim
31~\mathrm{km/s}$; when $\lambda_\odot=0.03~R_\odot$, we  do not 
find a steady-state
solution for $z_+^\odot \simeq 42~\mathrm{km/s}$. In Fig.~\ref{fig-1AU}  we
present
the values at 1 AU of wind  speed, density, mass flux, {temperature,
and heating per unit mass}
for the all the steady-state solutions. The different panels display how
the variations of the amplitude of the boundary condition $z_+^\odot$,
for  each value of  $\lambda_\odot$, 
changes the properties of the solution. Fig.~\ref{fig-1AU}a shows that an
increase in $z_+^\odot$  leads to higher densities at 1 AU, with less marked changes
in the wind speed. Not surprisingly, the same applies to the mass flux
 as a function of the wind speed (Fig.~\ref{fig-1AU}b). However, as 
 it appears from Fig.~\ref{fig-1AU}c, for a given $\lambda_\odot$, changing
$z_+^\odot$ has  noticeable effects both on the temperature and the density of the
wind at 1 AU. Figure~\ref{fig-1AU}d indicates that, by changing
the turbulent length scale $\lambda_\odot$, we increase both speed and
temperature of the wind at Earth.  
{This direct proportionality of wind temperature and density is
discussed in   the analysis of \citet{2012JGRA..117.9102E}, who showed
how the $T-V$ relationship, except for transient phenomena,
  is a particularly robust one. Finally, Figs.~\ref{fig-1AU}e and
\ref{fig-1AU}f
show the heating rate per unit mass as a function respectively of the temperature
and the wind speed. \citet{2007JGRA..112.7101V} presented solar wind
measurements and showed how the turbolent heating rates (both expected
and calculated) correlate with the
temperature and the wind speed. Notwithstanding the differences between
 the models, 
our results appear to be roughly within the spread in values
 shown in Figs.~7 and 10 of
\citet{2007JGRA..112.7101V}.}
If we  compare our values with those obtained from \emph{in situ} measurements
\citep[e.g.][]{2013SoPh..285..167S}, we find that selecting a
combination of $\lambda_\odot$ and
$z_+^\odot$ such as  $0.03~R_\odot \lesssim \lambda_\odot \lesssim 0.07~R_\odot $ 
and  $23~\mathrm{km/s} \lesssim z_+^\odot \lesssim 31~\mathrm{km/s} $ 
yields solutions compatible with  observations.

\subsection{Example of a Fast Wind Solution}
\label{sec-fast}
\begin{figure}
\includegraphics[width=\textwidth]{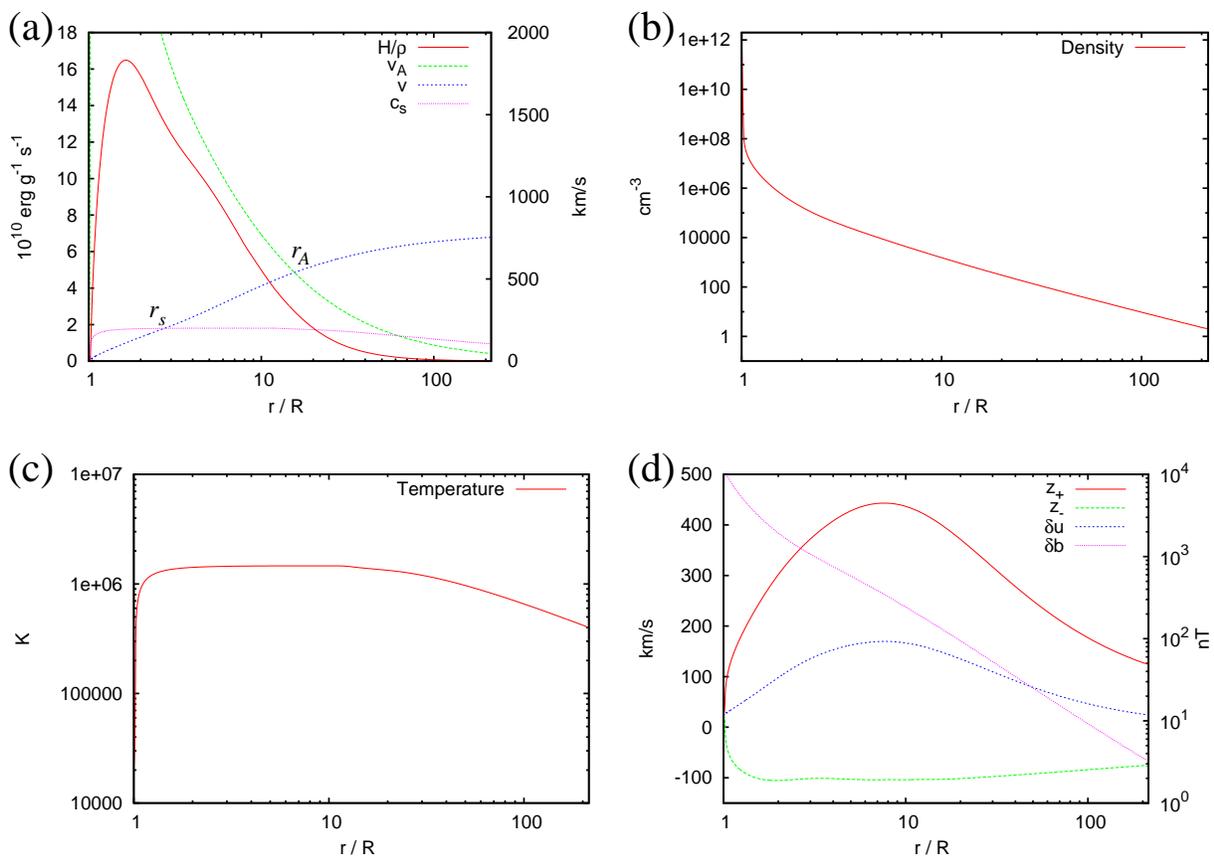}
\caption{Fast wind solution with $z_+^\odot=25~\mathrm{km/s} $, 
$\lambda_\odot=0.05~R_\odot$,
 thermal conduction, and radiative losses.  (a) Heating per unit mass (left
scale), wind speed, Alfv\'en speed, sound speed (right scale). The bulk of the heating is
deposited within the critical point for  Alfv\'en waves (marked $r_A$, while the
sound critical point is at 
$r_s$). (b) Number density; (e) Temperature. (d) The amplitudes of the
perturbations, $z_+$ and $z_-$, {of their average, $\delta u$ (left
scale), and of $\delta  b$ (right scale).} }
\label{fig-fast}
\end{figure}
We now discuss in more details  one of the solutions obtained in
Sec.~\ref{sec-thermo}. Among the possible choices of fast solar wind solutions,
we select the one with $z_+^\odot=25~\mathrm{km/s} $ and
$\lambda_\odot=0.05~R_\odot$, which yields a wind speed of 753 km/s and a
number 
density of {$2.06~\mathrm{cm}^{-3}$} at 1 AU.
In Fig.~\ref{fig-fast}a, we plot  wind speed,  Alfv\'en speed, and 
sound speed together with the heating per unit mass. The bulk of the heating is
deposited below the critical Alfv\'en point, where the wind is sub-Alfv\'enic.
Figures \ref{fig-fast}c and \ref{fig-fast}d show the behavior of the density and
temperature respectively. Introducing 
thermal conduction creates a 
temperature plateau between 2 and {10} $R_\odot$ and, consequently, also
a relatively constant sound speed (Fig.~\ref{fig-fast}a). In Fig.~\ref{fig-fast}c we 
plot the two components of the perturbation, $z_+$ and $z_-$,  their
average, $\delta u$, {and the perturbation of the magnetic field},
$\delta b$. The values of $\delta u$ are compatible with observations
that lie in the range $\delta u = 90 \pm 30~\mathrm{km/s}$ at $4~R_\odot$ and 
$\delta u = 150 \pm 100~\mathrm{km/s}$ at $10-15~R_\odot$
 \citep{1988ApJ...325..442W}.
{$\delta b$ is also 
within the range of values as measured
  by Helios 1 at a distance from the Sun between 0.66 and
0.60 AU \citep[Fig.~5-4 of][]{1995SSRv...73....1T} 
 and Helios 2  at a distance from the Sun of 0.9 AU
\citep[Fig.~9 of][]{2013LRSP...10....2B}.}

{
While the values of the photospheric correlation length we have used in the
model are
of the order of the supergranular sizes on the solar surface, there is evidence
from both direct observations of the photosphere \citep{2013ApJ...773..167A}
and radio scintillations of the corona \citep{2010ApJ...722.1495H} that
$\lambda_\odot$ may be as small as 0.0003 to 0.001 $R_\sun$. If, for the
case here described, we 
reduce the correlation length to 
$\lambda_\odot= 0.001~R_\sun$, this lowers
the wind speed at 1 AU to 610 km/s, but also  lowers the density
to only $0.7~\textrm{cm}^{-3}$, which is far below the measurements
of the slow solar wind.
Therefore it appears that smaller values
of $\lambda_\odot$ are not sufficient to reproduce the parameter space of
the slow solar wind, although the effects of the different properties
(e.g., larger expansion factors) 
of the magnetic field lines 
associated to the slow stream should be explored.
}
\section{DISCUSSION}
We have implemented a time-dependent model of the  solar wind with
 acceleration and heating through turbulence dissipation. Our model uses the
self-consistent formulation of \citet{2010ApJ...708L.116V} within            
a 1D HD code with thermal conduction and radiation losses.
For the case of a simplified energy equation, such as that used by
\citet{2010ApJ...708L.116V}, we accurately reproduce a solar wind solution.
When the model is extended 
by introducing thermal conduction
and radiative losses, our model produces
  fast solar wind solutions whose characteristics are compatible 
with \emph{in situ} measurements.
 The model of  \citet{2010ApJ...708L.116V} includes
a compressional heating in
the lower corona; we have found that it is not necessary to
include such phenomenological heating when a more realistic energy 
equation is used.
Our model, in comparison with that of \citet{2013AIPC.1539...30L},
is certainly more complicated, since it advances   the amplitude of
the perturbations rather than the  energies as the latter does. Moreover, the 
model of 
\citet{2013AIPC.1539...30L}
neglects the contribution of the Reynolds stress, assuming that outwardly
propagating wave is dominant. However, with the present method,
 we have found that the contribution of the 
Reynolds stress can be  sizable; in one particular case it lowered the
wind speed by approximately $25\%$.

Furthermore, if we consider the integration of turbulence dissipation heating and
acceleration in 3D MHD models, there are several advantages in the present formulation in respect
of that of  \citet{2013AIPC.1539...30L}. First,  it is not much more computationally 
demanding to advance the amplitudes rather the energies. Second, this
formulation in terms of $z_\pm$ does not require us to calculate  for each mesh
point the reflection coefficient along each field line passing through it.
{Third, in our experience 
the inclusion of time-dependent transport equations
for $z_\pm$ does not increase the physical  convergence time to steady state.}
Finally, as
shown by \citet{2011ApJ...727...84U}, the Reynolds stress term can be written
for a case when only perturbations perpendicular to the large scale magnetic
field are considered and whose actual directions do not play a significant
role.
Therefore we believe that the present formulation can be readily extended to
the 3D model
of \cite{2009ApJ...690..902L}.

\acknowledgements{ 
This work was supported by AFOSR,  NASA's LWS, HTP, and
strategic capabilities programs,
and by NSF through the Center for Integrated Space Weather Modeling (CISM).
This work was carried out in part by the Jet Propulsion Laboratory,
California Institute of Technology under a contract
with NASA. MV was supported by the NASA Solar Probe Plus
Observatory Scientist contract.
}

\bibliography{mybib}

\end{document}